# NUM-Based Rate Allocation for Streaming Traffic via Sequential Convex Programming

Ali Sehati
ECE Department, University of Tehran
School of Computer Science, IPM
Tehran, Iran.
Email: a.sehati@ipm.ir

Mohammad Sadegh Talebi
School of Computer Science, IPM
Tehran, Iran.
Email: mstalebi@ipm.ir

Ahmad Khonsari
ECE Department, University of Tehran
School of Computer Science, IPM
Tehran, Iran.
Email: khonsari@ut.ac.ir

*Abstract*—In recent years, there has been an increasing demand for ubiquitous streaming like applications in data networks. In this paper, we concentrate on NUM-based rate allocation for streaming applications with the so-called S-curve utility functions. Due to non-concavity of such utility functions, the underlying NUM problem would be non-convex for which dual methods might become quite useless. To tackle the non-convex problem, using elementary techniques we make the utility of the network concave, however this results in reverse-convex constraints which make the problem non-convex. To deal with such a transformed NUM, we leverage *Sequential Convex Programming (SCP)* approach to approximate the non-convex problem by a series of convex ones. Based on this approach, we propose a distributed rate allocation algorithm and demonstrate that under mild conditions, it converges to a locally optimal solution of the original NUM. Numerical results validate the effectiveness, in terms of tractable convergence of the proposed rate allocation algorithm.

## I. Introduction

With recent advances in networking technologies and video compression, there is an increasing demand for ubiquitous multimedia applications like live streaming, video gaming, video conferencing, and voice over IP. Multimedia applications are characterized by a multitude of QoS requirements including stringent bandwidth, delay, and delay jitter guarantees. The ever increasing demand for streaming traffic has attracted a lot of research interests to develop efficient mechanisms for resource allocation between competing multimedia sessions in a wide variety of networking scenarios [1]-[3].

In the course of the last decade, rate allocation has been widely addressed as the (usually distributed) solution to Network Utility Maximization (NUM), which has emerged as an analytical framework to understand and design existing network protocols [4]-[5]. The goal of NUM is to maximize the aggregate utility of the users subject to operational and practical constraints of the network. In the basic form of NUM proposed in [4], the feasibility of rate allocation was accommodated by congestion in links. So far, a plethora of studies have concentrated on NUM-based rate allocation for services with elastic traffic such as traditional file transfer. Due to strict concavity and differentiability of the utility function for elastic traffic, such NUMs are smooth and strictly convex and thus far have been efficiently solved using dual or primal-dual methods (see e.g. [5] and references therein.)

In contrast, applications that carry inelastic traffic like audio/video streaming, can only tolerate a limited amount of packet delay or fluctuation in rate. Hence, they are in possession of non-concave and often non-differentiable utility functions [1], [6]. This results in a non-convex and usually non-smooth NUM for which dual/primal-dual methods might prove quite useless.

There have been several works that have addressed non-convex NUM problems for resource allocation supporting inelastic services [6]-[12]. Lee *et al.* [7] outlined the possibility of divergence of dual methods for non-concave utilities and proposed a distributed "self-regulating" heuristic for rate control of non-concave utilities, where some of the sources turn themselves off according to their local information. Hande *et al.* [8] proposed necessary and sufficient conditions for canonical distributed algorithm to converge to global optimum in the presence of non-concave utilities. A centralized algorithm for non-convex NUM has been proposed in [9] in which sum-of-squares technique was applied to a polynomial approximation of the non-concave utility function. However, this centralized approach suffers from high order of complexity. In [6], the authors exerted a redefined variant of the non-concave utility function in a distributed flow control algorithm so that the network can achieve a utility-proportional fair rate allocation. Authors of [10] merged the utility-proportional theory with a stochastic optimization framework to propose a rate control algorithm for the mixture of elastic and inelastic traffic in wireless sensor networks. In [11], the authors introduced a smooth utility function as an approximation to the ideal staircase utility function for SVC-encoded streams and leveraged the utility-proportional approach to redefine the NUM which is solvable using dual methods. Authors of [12], addressed NUM problem in the context of random access in WLANs for stations generating either elastic or inelastic traffic.

In this study, we focus on NUM-based rate allocation for streaming applications with a class of non-concave utility functions. Towards this, we adopt the so-called *S-curve* utility functions for streaming traffic [2], [3] as they are shown to be capable of characterizing the user perceived quality for a broad range of multimedia streaming scenarios. In order to tackle the resulting non-convex NUM, we exploit transformation techniques to gain a strictly concave objective. However, this procedure yields a class of non-convex DC (difference of convex) constraints, referred to as *reverse-convex constraints* [13]. We then deal with the non-convex transformed NUM using an approach called *Sequential Convex Programming with DC constraints*, abbreviated as SCP-DC, which was proposed in [13]. In this regard, SCP-DC approach tackles the problem with reverse-convex constraints by solving a series of convex problems. Then we present a distributed rate allocation algorithm obtained by solving the sequence of convex problems in an iterative manner. We demonstrate that under mild assumptions, the proposed algorithm will converge to a locally optimal solution of the original NUM problem. To the best of our knowledge, this is the first work that addresses NUM with S-curve utilities with Sequential Convex Programming approach. Finally, our numerical experiments confirm the tractable convergence rate of our proposed algorithm and validate the its effectiveness in our experiment scenarios.

The rest of this paper is organized as follows. In Section II, we describe the network and utility model and in Section III, we establish problem formulation. Then we present our solution algorithm in Section IV. Numerical analysis is given in Section VI and conclusion is drawn in Section VII.

## II. SYSTEM MODEL

### A. Network Model

We consider a communication network that consists of a set $\mathcal{L} = \{1, \ldots, L\}$ of unidirectional links and a set $\mathcal{S} = \{1, \ldots, S\}$ of sources. We denote by $\mathbf{c} = (c_l, l \in \mathcal{L})$ the link capacity vector where $c_l$ is the capacity of link $l$ in bps. We assume that each logical source $s$ transmits at rate $x_s \in \mathcal{X}_s \triangleq [m_s, M_s]$, where $m_s$ and $M_s$ are the minimum and the maximum rates, respectively. There is a fixed set of links $\mathcal{L}(s) \subseteq \mathcal{L}$ that source $s$ uses to reach its destination. We represent such routes using a routing matrix $\mathbf{R} \in \mathbb{R}^{L \times S}$, which is defined as

$$R_{ls} = \begin{cases} 1 & \text{if source } s \text{ passes through link } l \\ 0 & \text{otherwise} \end{cases}$$

Rate allocation is considered to be feasible if and only if the source rate vector $\mathbf{x} = (x_s, s \in \mathcal{S})$ satisfies the following conditions

**C1.** $x_s \in \mathcal{X}_s, \quad s \in \mathcal{S}$
**C2.** $\sum_{s=1}^{S} R_{ls} x_s \leq c_l, \quad l \in \mathcal{L}$.

### B. Utility Model

In order to measure the user satisfaction degree, we use the well known notion of utility function. We associate an increasing and continuously differentiable function $U_s(x_s)$ with each source $s$. As mentioned in [1], multimedia applications, such as video streaming and VoIP, fall in the category of inelastic traffic and unlike elastic traffic, they are usually modeled by a family of non-concave utility functions referred to as sigmoidal-like functions[1] [7]. For example, previous studies have mostly used *sigmoidal logistic function* defined below as the utility function for inelastic traffic [7], [9]:

$$U(x) = \frac{1}{1 + e^{-\alpha(x-\beta)}} \quad (1)$$

which has the inflection point $x^{\text{infl}} = \beta$.

[1]An increasing function $f(x)$ is called a *sigmoidal-like function*, if it has one inflection point $x_0$, and $f''(x) > 0$, for $x < x_0$ and $f''(x) < 0$, for $x > x_0$. In other words, $f(x)$ is convex for $x < x_0$ and concave for $x > x_0$.

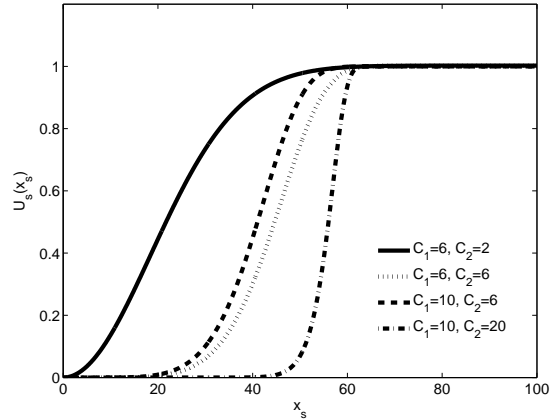

Fig. 1: S-curve Utility Corresponding to different values of $C_1$ and $C_2$

In this work, we focus on streaming applications which are shown to admit utility functions referred to as S-curve [3], [2]. Such utility functions can capture the perceptual video quality of streaming users as a function of transmission rate. In [2], the authors have proposed the following mathematical expression for this class of utility functions

$$U_s(x_s) = \frac{1 - e^{-C_{1s}\left(\frac{x_s}{r_s}\right)^{C_{2s}}}}{1 - e^{-C_{1s}}} \quad (2)$$

where $r_s$ is the constant rate at which video of source $s$ is encoded and $x_s$ is the average data rate received during transmission. Constants $C_{1s} > 0$ and $C_{2s} \geq 1$ are some parameters that depend on the properties of the video sequence and video encoder and might be determined in an offline fashion for stored media streaming.

It is easy to verify that the inflection point of the S-curve utility function is given by

$$x^{\text{infl}} = r_s \left(\frac{C_{2s} - 1}{C_{1s} C_{2s}}\right)^{\frac{1}{C_{2s}}}$$

The above equation implies that $C_{2s} > 1$ results in $x^{\text{infl}} > 0$ which makes S-curve a non-concave function. Fig. 1 portrays some utility functions corresponding to different values of parameters $C_{1s}$ and $C_{2s}$.

The family of S-curve utility functions are capable of capturing characteristics of fine-granularity scalable, layered and non-scalable video streams as their special cases. For example, perceptual quality of FGS encoded video is smooth and can easily be approximated by (2). Moreover, the step-wise utility of SVC video streams [14] can also be roughly characterized by S-curve (2). Hard real-time applications such as traditional voice service require fixed transmission rate. For such services, the utility below a threshold rate would be zero.

These applications are non-scalable and can be represented by a step utility function. S-curve utility function (2) can also approximate a step function as $C_{2s} \to \infty$ [2].

## III. PROBLEM FORMULATION

We model the rate allocation for streaming applications following the framework of Network Utility Maximization (NUM) which was proposed as the extension to optimization flow control in the seminal work of Low *et al.* [4]. The objective is the sum of utility functions with utilities defined by (2) and the constraints are feasibility conditions **C1**-**C2**. The rate allocation problem is described as follows

$$\max_{\boldsymbol{x} \in \mathcal{X}} \quad \sum_{s=1}^{S} U_s(x_s)$$
$$\text{subject to} \quad \sum_s R_{ls} x_s \leq c_l \quad \forall l \in \mathcal{L} \quad (3)$$

where $\mathcal{X}$ denotes the Cartesian product of all rate domains $\mathcal{X}_s, s \in \mathcal{S}$.

As stated in the previous section, the S-curve utility function (2) is non-concave for $C_{2s} > 1$ which makes the above problem non-convex. We would like to elaborate on making this problem convex so as to use powerful methods developed for convex optimization. We can make the utility function concave with the following change of variables:

$$\tilde{x}_s = \left(\frac{x_s}{r_s}\right)^{C_{2s}} \quad (4)$$

Substituting the above transformation in (2), we obtain the transformed utility function $\tilde{U}_s(.)$ as

$$\tilde{U}_s(\tilde{x}_s) = \frac{1 - e^{-C_{1s} \tilde{x}_s}}{1 - e^{-C_{1s}}} \quad (5)$$

where the transformed variable $\tilde{x}_s$ belongs to

$$\tilde{x}_s \in \tilde{\mathcal{X}}_s \triangleq \left[\left(\frac{m_s}{r_s}\right)^{C_{2s}}, \left(\frac{M_s}{r_s}\right)^{C_{2s}}\right]$$

The transformed utility function $\tilde{U}_s$ is strictly concave in $\tilde{x}_s$, because for $C_{1s} > 0$, its second derivative satisfies

$$\tilde{U}''_s(\tilde{x}_s) = -\frac{(C_{1s})^2}{1 - e^{-C_{1s}}} e^{-C_{1s}\tilde{x}_s} < 0 \quad (6)$$

Rewriting the capacity constraint for link $l$, yields

$$g_l(\tilde{\boldsymbol{x}}) \triangleq \sum_s R_{ls} r_s \tilde{x}_s^{\frac{1}{C_{2s}}} \leq c_l; \quad \forall l \in \mathcal{L}. \quad (7)$$

Unfortunately, the above capacity constraints with transformed variables do not correspond to a convex constraint as the (L.H.S) of (7) is a concave function. Indeed, the set $\mathcal{D}_l = \{\tilde{x}_s | g_l(\tilde{x}_s) - c_l \leq 0\}$ is a non-convex set, however the set $\mathbb{R}^S_+ - \mathcal{D}_l = \{\tilde{x}_s | g_l(\tilde{x}_s) - c_l > 0\}$ is a convex set. In optimization terminology, such a constraint is referred to as a *reverse-convex constraint* which is a special case of *Difference of Convex (DC) constraints* [13], [15].

In order to tackle such reverse-convex constraints, we use the *sequential convex programming algorithm with DC constraints (SCP-DC)* proposed in [13]. In this approach, the non-convex function that borders the range of permissible values for a constraint is replaced by an affine approximation to make the constraint convex. Using this approach, the L.H.S of each reverse-convex constraint $g_l(\tilde{\boldsymbol{x}}) \leq c_l$ is replaced by its first order Taylor approximation around a feasible point $\tilde{\boldsymbol{x}}'$, denoted by $\hat{g}_l(\tilde{\boldsymbol{x}}, \tilde{\boldsymbol{x}}')$, as follows

$$\hat{g}_l(\tilde{\boldsymbol{x}}, \tilde{\boldsymbol{x}}') \triangleq g_l(\tilde{\boldsymbol{x}}') + \nabla g_l(\tilde{\boldsymbol{x}}')^T (\tilde{\boldsymbol{x}} - \tilde{\boldsymbol{x}}') \leq c_l \quad (8)$$

Since $g_l$ is differentiable, $\nabla g$ exists at auxiliary variable $\tilde{x}'_s \in \tilde{\mathcal{X}}_s$. It's easy to verify that $\hat{g}_l(\tilde{\boldsymbol{x}}, \tilde{\boldsymbol{x}}')$ is affine in $\tilde{\boldsymbol{x}}$ and thereby L.H.S of (8) is convex. Thus, the constraint (8) represents a convex constraint. For $\hat{g}_l$ we get

$$\hat{g}_l(\tilde{\boldsymbol{x}}, \tilde{\boldsymbol{x}}') = \sum_s R_{ls} r_s \left((\tilde{x}'_s)^{\frac{1}{C_{2s}}} + \frac{1}{C_{2s}} (\tilde{x}'_s)^{\frac{1}{C_{2s}}-1} (\tilde{x}_s - \tilde{x}'_s)\right) \quad (9)$$

Finally, we rewrite the NUM problem with approximated constraints as

$$\max_{\tilde{\boldsymbol{x}}, \tilde{\boldsymbol{x}}' \in \tilde{\mathcal{X}}} \quad \sum_{s=1}^{S} \frac{1 - e^{-C_{1s}\tilde{x}_s}}{1 - e^{-C_{1s}}} \quad (10)$$
$$\text{subject to:} \quad \hat{g}_l(\tilde{\boldsymbol{x}}, \tilde{\boldsymbol{x}}') \leq c_l; \quad \forall l \in \mathcal{L}. \quad (11)$$

The above problem is strictly convex (in $\tilde{\boldsymbol{x}}$) since its objective is strictly concave because of (6) and its constraints are affine functions.

Before proceeding to solve the above problem, it's worth mentioning that in case of sigmoidal logistic utility functions (1), if we define $\tilde{x}_s = e^{\alpha(x-\beta)}$, we will come up with a convex objective with DC constraints, which can be treated by the aforementioned technique to obtain a convex formulation similar to problem (10)-(11). Therefore, the solution procedure to be discussed in the next section, will be applicable to the case of NUM with sigmoidal logistic utility functions.

## IV. OPTIMAL SOLUTION

In this section, we solve problem (10)-(11) using dual methods [4], [16].

### A. Primal Problem

The Lagrangian function is derived as [16]:

$$L(\tilde{\boldsymbol{x}}, \tilde{\boldsymbol{x}}', \boldsymbol{\mu}) = \sum_s \tilde{U}_s(\tilde{x}_s) - \sum_l \mu_l (\hat{g}_l(\tilde{\boldsymbol{x}}, \tilde{\boldsymbol{x}}') - c_l) \quad (12)$$

where $\mu_l$ is the positive Lagrange multiplier associated with constraint (11) for link $l$ and $\boldsymbol{\mu} = (\mu_l, l \in \mathcal{L})$.

According to Karush-Kuhn-Tucker (KKT) theorem, the stationary point of the Lagrangian, i.e. the solution to $\nabla L(\tilde{\boldsymbol{x}}^*, \tilde{\boldsymbol{x}}'^*, \boldsymbol{\mu}^*) = \boldsymbol{0}$, provides the unique solution to the problem (10)-(11). Partial derivatives of the Lagrangian with respect to $\tilde{x}_s$ is given in (13) at the top of the next page, and finally for the stationary point we get

$$\tilde{x}^*_s = \frac{1}{C_{1s}} (A^*_s - \log \rho^{s*}) \quad (14)$$

where

$$A^*_s = \log \left(\frac{C_{1s} C_{2s}}{r_s (1 - e^{-C_{1s}})}\right) + \left(1 - \frac{1}{C_{2s}}\right) \log \tilde{x}'^*_s$$
$$\rho^{s*} = \sum_l R_{ls} \mu^*_l. \quad (15)$$

$$\begin{aligned}\frac{\partial L}{\partial \tilde{x}_s} &= \frac{d}{d\tilde{x}_s}\tilde{U}_s(\tilde{x}_s) - \frac{d}{d\tilde{x}_s}\sum_l \mu_l \left(\sum_s R_{ls}r_s\left((\tilde{x}'_s)^{\frac{1}{C_{2s}}} + \frac{1}{C_{2s}}(\tilde{x}'_s)^{\frac{1}{C_{2s}}-1}(\tilde{x}_s - \tilde{x}'_s)\right) - c_l\right) \\ &= \frac{C_{1s}e^{-C_{1s}\tilde{x}_s}}{1-e^{-C_{1s}}} - \frac{d}{d\tilde{x}_s}\left((\tilde{x}'_s)^{\frac{1}{C_{2s}}} + \frac{1}{C_{2s}}(\tilde{x}'_s)^{\frac{1}{C_{2s}}-1}(\tilde{x}_s - \tilde{x}'_s)\right)r_s\sum_l R_{ls}\mu_l \\ &= \frac{C_{1s}e^{-C_{1s}\tilde{x}_s}}{1-e^{-C_{1s}}} - \frac{r_s}{C_{2s}}(\tilde{x}'_s)^{\frac{1}{C_{2s}}-1}\sum_l R_{ls}\mu_l = 0 \end{aligned} \quad (13)$$

It's easy to verify that the transformed source rate $\tilde{x}_s^*$ is a decreasing function with respect to $\mu_l^*, l \in \mathcal{L}$.

We postpone finding $\tilde{x}'^*$ to the next subsection. We will find $\boldsymbol{\mu}^*$ by solving the dual problem associated to the primal problem. Towards this, we first derive the dual function, which is defined as the following Lagrangian maximization [16]:

$$D(\boldsymbol{\mu}) = \max_{\tilde{\boldsymbol{x}}, \tilde{\boldsymbol{x}}' \in \mathcal{X}} L(\tilde{\boldsymbol{x}}, \tilde{\boldsymbol{x}}', \boldsymbol{\mu}) = L(\tilde{\boldsymbol{x}}^*, \tilde{\boldsymbol{x}}'^*, \boldsymbol{\mu})$$

### B. Dual Problem

Having obtained the dual function, i.e. $D(\boldsymbol{\mu}) = L(\tilde{\boldsymbol{x}}^*, \tilde{\boldsymbol{x}}'^*, \boldsymbol{\mu})$, the dual problem is defined as the following minimization problem [16]:

$$\min_{\boldsymbol{\mu} \geq \mathbf{0}} D(\boldsymbol{\mu}) \quad (16)$$

Solving the above problem in closed form might be impossible, and hence we solve it using iterative methods. As problem (10)-(11) is strictly convex, the dual function $D(\boldsymbol{\mu})$ is differentiable over the open set $\mathbb{R}_{++}^L$ and we can benefit from *gradient projection algorithm* to solve the dual problem [17]. In this algorithm, the dual variable is iteratively updated in the opposite direction to $\nabla D(\boldsymbol{\mu})$ as follows:

$$\boldsymbol{\mu}^{(t+1)} = [\boldsymbol{\mu}^{(t)} - \gamma \nabla D(\boldsymbol{\mu}^{(t)})]^+$$

where $\gamma > 0$ is a sufficiently small step-size.

Using Danskin's Theorem [17], the partial derivatives of the dual function are characterized as follows

$$\frac{\partial D}{\partial \mu_l} = c_l - \hat{g}_l(\tilde{\boldsymbol{x}}, \tilde{\boldsymbol{x}}') \quad (17)$$

In an iterative setting, to find the optimal value of the auxiliary variable $\tilde{\boldsymbol{x}}'^*$, similar to Proximal Optimization Methods [17], we update it as follows

$$\tilde{\boldsymbol{x}}'^{(t+1)} = \tilde{\boldsymbol{x}}^{(t)}$$

Put another way, it makes sense that, to calculate primal-optimal variable $\tilde{\boldsymbol{x}}^{(t+1)}$ at iteration step $t$, $\tilde{\boldsymbol{x}}^{(t)}$ is the best candidate for $\tilde{\boldsymbol{x}}'$ along which the affine approximation (8) can be made.

Substituting (17) into gradient projection update formula results in the following dual variable update

$$\begin{aligned}\mu_l^{(t+1)} &= \left[\mu_l^{(t)} - \gamma(c_l - \hat{g}_l(\tilde{\boldsymbol{x}}^{(t)}, \tilde{\boldsymbol{x}}'^{(t)}))\right]^+ \\ &= \left[\mu_l^{(t)} - \gamma(c_l - \hat{g}_l(\tilde{\boldsymbol{x}}^{(t)}, \tilde{\boldsymbol{x}}^{(t-1)}))\right]^+ \quad (18)\end{aligned}$$

where $\tilde{\boldsymbol{x}}^{(t)}$ is the value of optimal transformed rate given $\boldsymbol{\mu}^{(t)}$.

Moreover, for rate computation at iteration $t$, we get

$$\tilde{x}_s^{(t+1)} = \frac{1}{C_{1s}}\left(A_s^{(t+1)} - \log \rho^{s(t)}\right) \quad (19)$$

where

$$A_s^{(t+1)} = \log\left(\frac{C_{1s}C_{2s}}{r_s(1-e^{-C_{1s}})}\right) + \left(1 - \frac{1}{C_{2s}}\right)\log \tilde{x}_s^{(t)}$$

Finally, by taking the inverse transformation of (4), for source rate at iterate $t+1$, we get

$$x_s^{(t+1)} = \left[r_s\left(\tilde{x}_s^{(t+1)}\right)^{\frac{1}{C_{2s}}}\right]_{\mathcal{X}_s} \quad (20)$$

where $[.]_{\mathcal{X}_s}$ is the projection onto $\mathcal{X}_s$.

## V. Rate Allocation Algorithm

### A. Algorithm

The equations obtained in the previous section for optimal source rate calculation, i.e. (19) and (20), and dual variable update, i.e. (18), can work together to form a distributed solution to problem (3). Below we have shown a concise form of this iterative algorithm as Algorithm 1. As we can see, implementation of this algorithm necessitates two mechanisms for information exchange between links and sources.

1) Each link $l$ updates its price and communicates the result to the corresponding sources.
2) Each source $s$ calculates its new rate and informs the links in its path.

This type of information exchange can be carried out explicitly, for example via flooding-like mechanisms as suggested in [18]. This is in contrast to the implicit mechanisms inherent in Optimization Flow Control approach [4] where each source can infer aggregate price of its path using either queueing delay or packet loss ratio, and each link just needs to measure its current flow to update its price.

### B. Convergence

First we note that at the steady state, i.e. when

$$\tilde{\boldsymbol{x}}^{(t+1)} = \tilde{\boldsymbol{x}}^{(t)},$$

or equivalently when

$$\tilde{\boldsymbol{x}}'^{(t+1)} - \tilde{\boldsymbol{x}}^{(t+1)} = 0,$$

the approximated capacity constraints (8) would become

$$\sum_s R_{ls}r_s(\tilde{x}_s^{(t+1)})^{\frac{1}{C_{2s}}} \leq c_l, \quad l \in \mathcal{L}, \quad (21)$$

and thereby convexified constraints (8) will become equivalent to DC constraints (7). Therefore, if the algorithm converges to the steady state, the convexified constraints (8) will be equivalent to the original constraints of the transformed problem. In [13], it has been proved that under mild conditions on the objective function, such as strict convexity, the SCP-DC algorithm converges to a local maximizer of the non-convex problem (3). Therefore, the proposed rate allocation algorithm will reach a local maximum of problem (3) provided that $\gamma$ is chosen sufficiently small so that the gradient projection algorithm will converge [17].

Based on the results stated in [13], if SCP-DC algorithm converges, then the steady state point is a stationary point of the optimization problem. Put another way, the steady state point is a local optimal of the optimization problem. It has also been proved in [13] that for strictly convex objectives, the SCP-DC algorithm always converges to a KKT point of the original optimization problem (3).

---

**Algorithm 1**
**Distributed Rate Control Algorithm for Streaming Traffic Using SCP-DC Algorithm**

**Initialization**
1) Set of sources and links including the routing matrix
2) $r_s, C_{1s}, C_{2s}$ for $s \in \mathcal{S}$
3) $\gamma$ and $c_l$ for $l \in \mathcal{L}$

**Main Loop**
Do until $\max_s |x_s^{(t+1)} - x_s^{(t)}| < \epsilon$

1) For each $l \in \mathcal{L}$, update link price $\mu_l$ by:
$$\mu_l^{(t+1)} = \left[\mu_l^{(t)} - \gamma(c_l - \hat{g}_l(\tilde{\boldsymbol{x}}^{(t)}, \tilde{\boldsymbol{x}}^{(t-1)}))\right]^+$$

2) For each $s \in \mathcal{S}$, $\tilde{x}_s$ is calculated by:
$$A_s^{(t+1)} = \log\left(\frac{C_{1s}C_{2s}}{r_s(1-e^{-C_{1s}})}\right) + \left(1 - \frac{1}{C_{2s}}\right)\log \tilde{x}_s^{(t)}$$
$$\tilde{x}_s^{(t+1)} = \frac{1}{C_{1s}}\left(A_s^{(t+1)} - \log \rho^{s(t)}\right)$$
and $\rho^{s(t)} = \sum_l R_{ls}\mu_l^{(t)}$. Then calculate $x_s^{(t+1)}$ as
$$x_s^{(t+1)} = \left[r_s\left(\tilde{x}_s^{(t+1)}\right)^{\frac{1}{C_{2s}}}\right]_{\mathcal{X}_s}$$

---

## VI. NUMERICAL ANALYSIS

In this section, we investigate the performance and validity of the proposed rate allocation algorithm listed in the previous section as Algorithm 1.

### A. Scenario 1

We first consider a simple topology with a single bottleneck link with capacity $c = 1$ Mbps. Video sequences for all sources are assumed to be encoded at the constant bit rate $r_s = 256$ Kbps. Sources have utility functions with parameters $C_{1s} = 6, \forall s$ and $(C_{21}, \ldots, C_{25}) = (2, 4, 6, 8, 10)$. The 2nd column of Table I lists the results of the proposed rate allocation algorithm with step size $\gamma = 10^{-4}$ and the stopping criterion $\epsilon = 0.1$.

Fig. 3(a) and Fig. 3(b) display the evolution of source rates and link price, respectively. From Fig. 3(a), it is observable that the stopping criterion is met in iteration step $t = 12$ which implies the proper convergence rate of the algorithm.

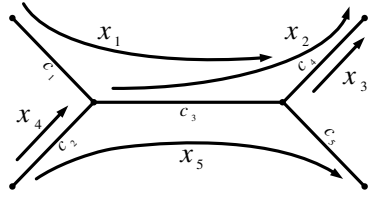

Fig. 2: Network topology and flow rates

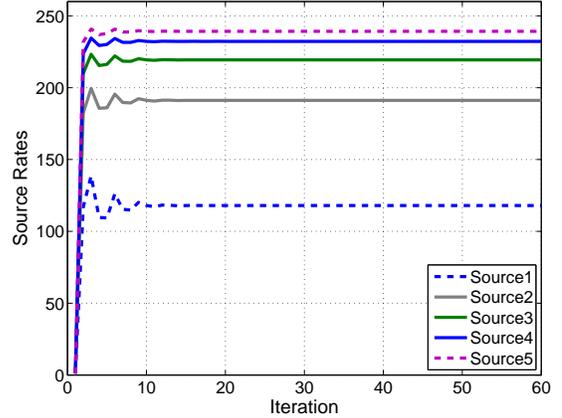

(a) Evolution of Flow Rates

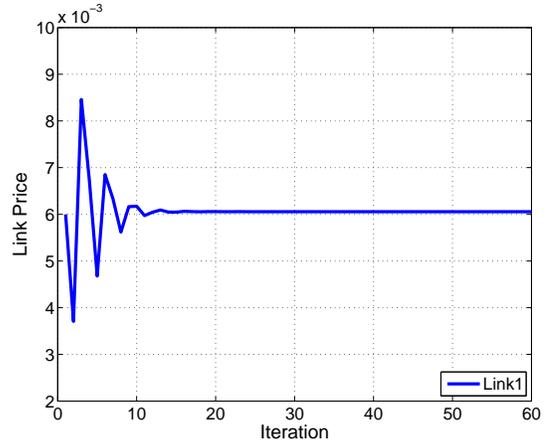

(b) Evolution of Shadow Price

Fig. 3: Evolution of (a) Source rates and (b) Link price for the first scenario

### B. Scenario 2

Now we focus on a scenario with multiple bottleneck links whose topology is shown in Fig. 2 with capacity vector $\boldsymbol{c} = (210, 425, 610, 425, 210)$ Kbps. Sources have utility functions with parameters $C_{2s} = 6, \forall s$ and $(C_{11}, \ldots, C_{15}) = (2, 4, 6, 8, 10)$. Similar to the previous scenario, we set $r_s = 256$, $\gamma = 10^{-4}$, and $\epsilon = 0.1$. The 4th column of Table I lists the results of the proposed rate allocation algorithm. Fig. 4(a) and Fig. 4(b) display the evolution of source rates and link prices, respectively. As shown in Fig. 4(a), the stopping criterion is satisfied in iteration step $t = 70$ which again demonstrates the tractable convergence rate of the proposed algorithm in a topology with multiple bottleneck links.

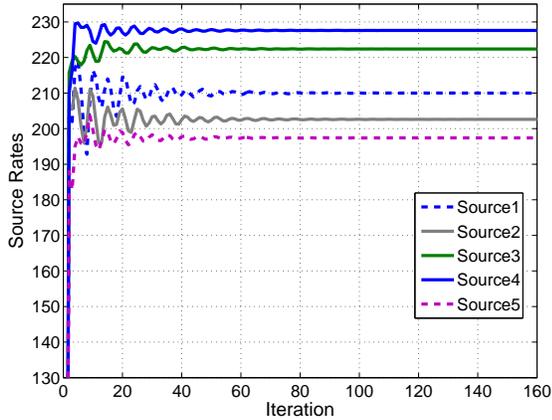

(a) Evolution of Flow Rates

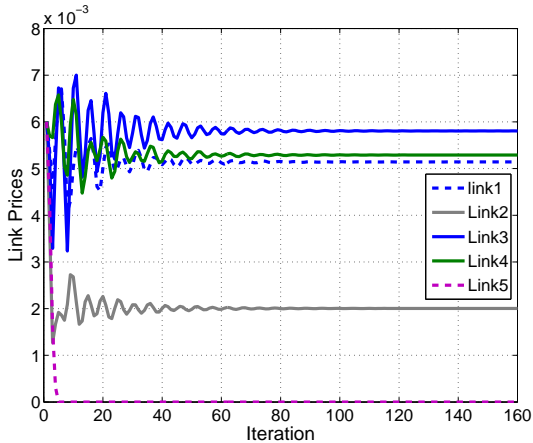

(b) Evolution of Shadow Prices

Fig. 4: Evolution of (a) Source rates and (b) Link prices for the second scenario

|  | Scenario 1 | | Scenario 2 | |
|---|---|---|---|---|
| **Source** | Algorithm | fmincon | Algorithm | fmincon |
| 1 | 118.1668 | 117.9658 | 209.8065 | 210.0000 |
| 2 | 191.2863 | 191.1745 | 202.4938 | 202.6043 |
| 3 | 219.4306 | 219.3638 | 222.3787 | 227.3957 |
| 4 | 232.2970 | 232.2520 | 227.6564 | 227.6043 |
| 5 | 239.2770 | 239.2439 | 197.3906 | 197.3957 |

TABLE I: Rate allocation results

## C. Validation

In order to validate the rate allocation results obtained above, we have also solved the problem (3) by invoking `fmincon` function in Matlab [19]. When calling this function, we choose *Interior-Point Method* [16], [17] as its solving algorithm. The results returned by `fmincon` for the two scenarios along with those obtained by our algorithm are listed in Table I. It is easy to confirm that the rate allocation results completely match those obtained from `fmincon`.

## VII. CONCLUSION

In this paper, we addressed rate allocation for streaming applications with non-convex S-curve utility functions. First we convexified the utility functions with elementary transformation techniques. Then, we exploited the SCP-DC approach [13] to handle the resultant reverse-convex constraints. Using dual method, we then proposed a distributed rate allocation algorithm which was shown to achieve a locally optimal solution of the non-convex NUM. Simulation results validated the tractable convergence and accuracy of the proposed rate allocation algorithm. As a possible direction to continue this research, it is promising to address rate allocation for such streaming applications in wireless networks.